\documentclass[letterpaper,twocolumn,prl,aps,superscriptaddress,amsmath,amssymb,floatfix]{revtex4-1}
\usepackage{mathptmx}
\usepackage[latin9]{inputenc}
\usepackage{color}
\usepackage{amsmath}
\usepackage{amssymb}
\usepackage{graphicx}
\usepackage{esint}
\usepackage[unicode=true,
 bookmarks=true,bookmarksnumbered=false,bookmarksopen=false,
 breaklinks=false,pdfborder={0 0 1},backref=false,colorlinks=true]
 {hyperref}
\hypersetup{
 linkcolor=magenta,urlcolor=blue,citecolor=blue,pdfstartview={FitH},hyperfootnotes=false}

\makeatletter

\usepackage{textcomp}
\usepackage{epstopdf}
\usepackage{amsfonts}
\usepackage{xcolor}
\usepackage{soul}
\makeatother

\begin{document}
\title{Multicolor continuous-variable quantum entanglement in the Kerr frequency
comb}
\author{Ming Li}
\affiliation{Key Laboratory of Quantum Information, CAS, University of Science
and Technology of China, Hefei 230026, China}
\affiliation{CAS Center For Excellence in Quantum Information and Quantum Physics,
University of Science and Technology of China, Hefei, Anhui 230026,
P. R. China.}
\author{Yan-Lei Zhang}
\affiliation{Key Laboratory of Quantum Information, CAS, University of Science
and Technology of China, Hefei 230026, China}
\affiliation{CAS Center For Excellence in Quantum Information and Quantum Physics,
University of Science and Technology of China, Hefei, Anhui 230026,
P. R. China.}
\author{Xin-Biao Xu}
\affiliation{Key Laboratory of Quantum Information, CAS, University of Science
and Technology of China, Hefei 230026, China}
\affiliation{CAS Center For Excellence in Quantum Information and Quantum Physics,
University of Science and Technology of China, Hefei, Anhui 230026,
P. R. China.}
\author{Chun-Hua Dong}
\affiliation{Key Laboratory of Quantum Information, CAS, University of Science
and Technology of China, Hefei 230026, China}
\affiliation{CAS Center For Excellence in Quantum Information and Quantum Physics,
University of Science and Technology of China, Hefei, Anhui 230026,
P. R. China.}
\affiliation{Hefei National Laboratory, University of Science and Technology of China, Hefei 230088, China.}

\author{Guang-Can Guo}
\affiliation{Key Laboratory of Quantum Information, CAS, University of Science
and Technology of China, Hefei 230026, China}
\affiliation{CAS Center For Excellence in Quantum Information and Quantum Physics,
University of Science and Technology of China, Hefei, Anhui 230026,
P. R. China.}
\affiliation{Hefei National Laboratory, University of Science and Technology of China, Hefei 230088, China.}

\author{Chang-Ling Zou}
\email{clzou321@ustc.edu.cn}
\affiliation{Key Laboratory of Quantum Information, CAS, University of Science
and Technology of China, Hefei 230026, China}
\affiliation{CAS Center For Excellence in Quantum Information and Quantum Physics,
University of Science and Technology of China, Hefei, Anhui 230026,
P. R. China.}
\affiliation{Hefei National Laboratory, University of Science and Technology of China, Hefei 230088, China.}

\date{\today}

\begin{abstract}
In a traveling wave microresonator, the cascaded four-wave mixing (FWM) between optical modes allows the generation of frequency combs, including intriguing dissipative Kerr solitons (DKS). In this study, we explore the quantum fluctuations of frequency combs and unveil the quantum features of solitons. The entanglement of DKS exhibits two distinct characteristics. For modes located at the spectral edge with a small number of excitations, different comb modes with multiple colors become entangled due to photon-pair generation and coherent photon conversion stimulated by the FWM. Notably, we observe a sudden disappearance of quantum entanglement in the center of the DKS spectrum, which is attributed to the self-locking phenomena in the FWM network. Our findings demonstrate the prominent quantum nature of DKS, which is of fundamental significance in quantum optics and has the potential to be utilized in quantum networking and distributed quantum sensing applications.
\end{abstract}
\maketitle

\section{Introduction}
In recent decades, nonlinear optics technologies have become the backbone of modern optics, allowing for frequency conversion, optical parametric oscillation, ultrafast modulation, and non-classical quantum sources. One of the most exciting advances is the optical frequency comb, which has attracted tremendous research interest due to its revolutionary applications in precision spectroscopy, astronomy detection, optical clocks, and optical communication~\citep{fortier201920,foltynowicz2011optical,picque2019frequency,Papp:14,obrzud2019microphotonic,marin2017microresonator,menicucci2008one,yao2022soliton,Sun2023}.
By harnessing the enhanced nonlinear optical effect in a microresonator~\cite{Liu2022}, the Kerr frequency comb, in particular, the DKS as a phase-locked frequency comb, has been extensively studied~\citep{kippenberg2018dissipative}. These DKS have been realized on photonic chips with various $\chi^{(3)}$ materials, demonstrating significant progress in compact size, scalability, low power consumption, great spectral range and large repetition rate~\citep{gaeta2019photonic,Wan:20,Bruch2021,bai2020brillouin,niu2023khz,ye2023foundry,liu2022optical}.

The DKS is appealing for photonic quantum information science. From one aspect, the DKS produces an array of stable and locked coherent laser sources with equally spaced frequencies, which provides a coherent laser source for driving nonlinear light-matter interaction as well as a stable frequency reference of local oscillators for heterodyne measurements via frequency multiplexing. By taking advantage of the coherence over a large frequency band, the quantum key distribution utilizing DKS has been experimentally studied~\citep{wang2020quantum}, which promises commercialized high-speed quantum communication. From another aspect, the Kerr nonlinear interaction is inherently a quantum parametric process that describes the annihilation of two photons and simultaneous generation of signal-idler photon pairs, which imply interesting quantum statistics of comb lines~\citep{chembo2016quantum,guidry2022quantum}. When working below the threshold, DKS devices have been exploited to generate both discrete- and continuous-variable (CV) quantum entangled states~\citep{reimer2016generation,Cui2020,zhu2020chip,PhysRevResearch.2.023138}, holding the potential for one-way quantum computing and high-dimensional entanglement between different colors that can be distributed over the quantum network~\citep{kues2019quantum}. For the case above the threshold, although previous studies of $\chi^{(2)}$ optical combs indicate the potential applications of combs in multipartite cluster-state generation for quantum computing~\citep{menicucci2008one,PhysRevLett.107.030505,PhysRevLett.108.083601}, the quantum entanglement of DKS has not yet been systemically studied.

In this work, we theoretically investigated the CV entanglement between comb lines in a microcavity and compared the quantum features of combs in different stable states. We showed that the DKS can stimulate pair-generation and coherent conversion interactions between optical modes and thus produces a complex network linking many optical modes. Evaluated by the entanglement logarithm negativity, it is demonstrated that a large number of modes located at the edge of the DKS spectra are entangled together. This multicolor quantum-entangled state~\citep{coelho2009three} holds great potential in applications in multi-user quantum communication, quantum teleportation networks~\citep{van2000multipartite,furusawa2007quantum}, and quantum-enhanced measurements~\citep{PhysRevLett.111.093603,simon2017quantum,guo2020distributed,xia2020demonstration}. However, for comb lines situated at the spectral center, comb frequencies are self-organized into a highly precise and equally spaced frequency layout. This process of phase locking effectively minimizes the fluctuations of the comb field, leading to the absence of CV entanglement. The phase transition of entanglement is elucidated by a simple frequency locking model that consists of only two FWM processes.

\section{Model and principle}
A typical device for generating a Kerr frequency comb consists of an optical microring cavity sided-coupled to a waveguide ~\citep{kippenberg2018dissipative}. By injecting a pump laser into a resonance of the cavity, the intracavity optical field builds up, and the photons are converted between optical modes via cascaded four-wave mixing (FWM) due to the Kerr nonlinearity, leading to a broad comb spectrum. To investigate the DKS in the microcavity, we consider a group of $2N+1$ modes belonging to the same mode family, with the spatial distributions described by $\Psi\left(\overrightarrow{r},\theta\right)=\phi\left(\overrightarrow{r}\right)e^{im\theta}$~\citep{strekalov2016nonlinear}. Here, $m\in\mathbf{m}$ denotes the model index, which corresponds to the angular momentum of the resonant modes and $\mathbf{m}=\left\{ -N,-N-1,...,-1,0,1,...,N-1,N\right\}$. Due to the dispersion, the resonant frequencies can be expanded with respect to the index as $\omega_{m}=\omega_{0}+mD_{1}+\frac{m^{2}}{2!}D_{2}+....$~\citep{chembo2013spatiotemporal}. The Hamiltonian describing this multimode system reads~\citep{guo2018efficient}
\begin{eqnarray}
H & = & \sum_{j=-N}^{N}\Delta_{i}a_{i}^{\dagger}a_{i}+g_{0}\sum_{ijkl}\delta\left(i+j-k-l\right)a_{i}^{\dagger}a_{j}^{\dagger}a_{k}a_{l}\label{eq:TotalH-1}
\end{eqnarray}
in the rotating frame of $\sum_{j=-N}^{N}\left(\omega_{p}+jD_{1}\right)a_{j}^{\dagger}a_{j}$. Here, $\Delta_{j}=\omega_{0}-\omega_{p}+j^{2}D_{2}/2$ is the mode detuning by neglecting higher order dispersion terms, $a_{i}^{\dagger}$($a_{i}$) is the photon creation (annihilation) operator, $g_{0}$ is the vacuum coupling strength of the Kerr nonlinearity, $\delta(\cdot)$ is the Kronecker delta function and reflects the phase-matching condition $i+j=k+l$ according to angular momentum conservation~\citep{guo2018efficient,strekalov2016nonlinear}. The nonlinear interaction terms describe the simultaneous annihilation and creation of photon-pairs in mode-pairs $\left(i,j\right)$ and $\left(k,l\right)$, with the total photon numbers conserved. By taking all permutations of indices $\left\{ i,j,k,l\right\}$ in the summation, the Hamiltonian is Hermitian and involves all FWM terms, including self-phase modulation, cross-phase modulation, and degenerate and non-degenerate FWMs.

The complex network of FWM in a microcavity implies complicated dynamics of optical fields, which can be numerically evaluated by solving the quantum dynamics of bosonic modes according to the Hamiltonian {[}Eq. (\ref{eq:TotalH-1}){]}. Instead of solving the intractable many-body nonlinear equations via full quantum theory, we adopt the mean-field treatment of the strongly pumped system, i.e., the mean and fluctuation of the optical field in each mode are solved separately, due to the weak nonlinearity $g_{0}/\kappa\ll1$ in practice~\citep{PhysRevApplied.13.034030}. The operator of the cavity mode $a_{i}$ is approximated by the sum
of a classical field described by an \emph{amplitude} $\alpha_{i}$
and a \emph{fluctuation} described by a bosonic operator $\delta a_{i}$.
According to the Heisenberg equation and discarding the high-order
terms of the fluctuation operators, the dynamics of the classical
fields and their fluctuations follow
\begin{eqnarray}
\frac{d}{dt}\alpha_{i} & = & \beta_{i}\alpha_{i}-2ig_{0}\sum_{jkl}\alpha_{j}^{*}\alpha_{k}\alpha_{l}+\varepsilon_{p}\delta\left(i\right),\label{eq:OperatorDm}\\
\frac{d}{dt}\delta a_{i} & = & \beta_{i}\delta a_{i}+\sqrt{2\kappa_{a}}a_{i}^{in}\nonumber \\
 &  & -2ig_{0}\sum_{jkl}\left(\alpha_{k}\alpha_{l}\delta a_{j}^{\dagger}+\alpha_{j}^{*}\alpha_{l}\delta a_{k}+\alpha_{j}^{*}\alpha_{k}\delta a_{l}\right),\label{eq:flucDm}
\end{eqnarray}
respectively. Here, the summation takes over all possible permutations $\left(j,k,l\right)$ with $i=k+l-j$, $\beta_{i}=-i\delta_{i}-\kappa_{i}$, $\kappa_{j}$ is the total amplitude decay rate of the $j$-th mode, $\varepsilon_{p}$ is the pump strength, and $a_{i}^{in}$ is the input noise on mode $i$ and fulfills $\langle a_{i}^{in}(t)(a_{i}^{in})^{\dagger}(t')\rangle=\delta(t-t')$. From Eq.~(\ref{eq:flucDm}), the last three terms represent the classical-field-stimulated photon-pair generation in modes $\left(i,j\right)$ and the coherent photon conversion between modes $\left(i,k\right)$ and $\left(i,l\right)$. Rather than merely produce downconversion photons by drive lasers~\citep{reimer2016generation}, these terms imply very rich quantum dynamics in the complex network, with the quantum correlations  between optical modes being directly generated through photon-pair generation and indirectly generated by coherently redistributing the generated photons among the modes.

The classical fields have been extensively studied in previous works~\citep{hansson2014numerical,guo2018efficient}. With the amplitudes of classical fields $\alpha_{i}$ obtained, the CV quantum correlation~\citep{braunstein2005quantum} between the comb lines can be evaluated by solving the dynamics of fluctuations {[}Eq.~(\ref{eq:flucDm}){]}. It is convenient to represent the fluctuations by the ``amplitude'' and ``phase'' field quadratures $\delta X_{i}=\left(\delta a^{\dagger}+\delta a\right)/\sqrt{2}$, $\delta Y_{i}=i\left(\delta a^{\dagger}-\delta a\right)/\sqrt{2}$. The dynamics of the quadratures $\overrightarrow{Q}=\left\{ \delta X_{-N},\delta Y_{-N},\delta X_{-N+1},\delta Y_{-N+1},...,\delta X_{N},\delta Y_{N}\right\} ^{T}$ follow $\frac{d}{dt}\overrightarrow{Q}=\mathbf{M}\cdot\overrightarrow{Q}+\overrightarrow{n}(t)$, where $\overrightarrow{n}^{T}=\left\{ \sqrt{2\kappa_{-N}}X_{-N}^{in},\sqrt{2\kappa_{-N}}Y_{-N}^{in},...\right\}$ is the input noise, and $\mathbf{M}$ is a $(4N+2)\times(4N+2)$ matrix derived from Eq.~(\ref{eq:flucDm}). Under the stability condition by the Routh-Hurwitz criterion~\citep{PhysRevA.35.5288}, the correlation matrix $\mathbf{V}$ for all modes can be solved following the deviation in Ref.~\citep{vitali2007optomechanical} and we obtain
\begin{eqnarray}
\mathbf{M}\mathbf{V}+\mathbf{V}\mathbf{M}^{T} & = & -\mathbf{D},\label{eq:corr}
\end{eqnarray}
where the element of the correlation matrix $V_{ij}=\langle Q_{i}Q_{j}+Q_{j}Q_{i}\rangle/2$ and
the noise term $D_{ij}=\langle n_{i}n_{j}+n_{j}n_{i}\rangle/2$ can
be derived from $\langle X_{i}^{in}(t)X_{j}^{in}(t')\rangle=\delta(i-j)\delta(t-t')$
and $\langle X_{i}^{in}(t)Y_{j}^{in}(t')\rangle=0$. The bipartite
CV entanglement between comb lines can be evaluated by the logarithmic
negativity
\begin{eqnarray}
E_{N}^{mn} & = & \mathrm{max}[0,-\ln\sqrt{2}\eta],\label{eq:En}
\end{eqnarray}
where $\eta=\sqrt{\Theta-\sqrt{\Theta^{2}-4\mathrm{det}V}}$, $\Theta=\mathrm{det}A+\mathrm{det}B-2\mathrm{det}C$,
with $A$, $B$ and $C$ are the elements of $V^{mn}=\{\{A,C\},\{C^{T},B\}\}$,
which is a sub-matrix of $V$ representing the bipartite correlation
matrix between $\left\{ \delta X_{m},\delta Y_{m},\delta X_{n},\delta Y_{n}\right\} $.
$E_{N}^{mn}$ is a measure of the CV entanglement~\citep{PhysRevA.70.022318,PhysRevA.65.032314,vitali2007optomechanical} and the mode-pair is entangled only if $E_{N}^{mn}>0$. By calculating all combinations of the modes, one obtains the entanglement matrix
$\mathbf{E}_{N}$ of the cavity field.

\begin{figure}
\begin{centering}
\includegraphics[width=1\linewidth]{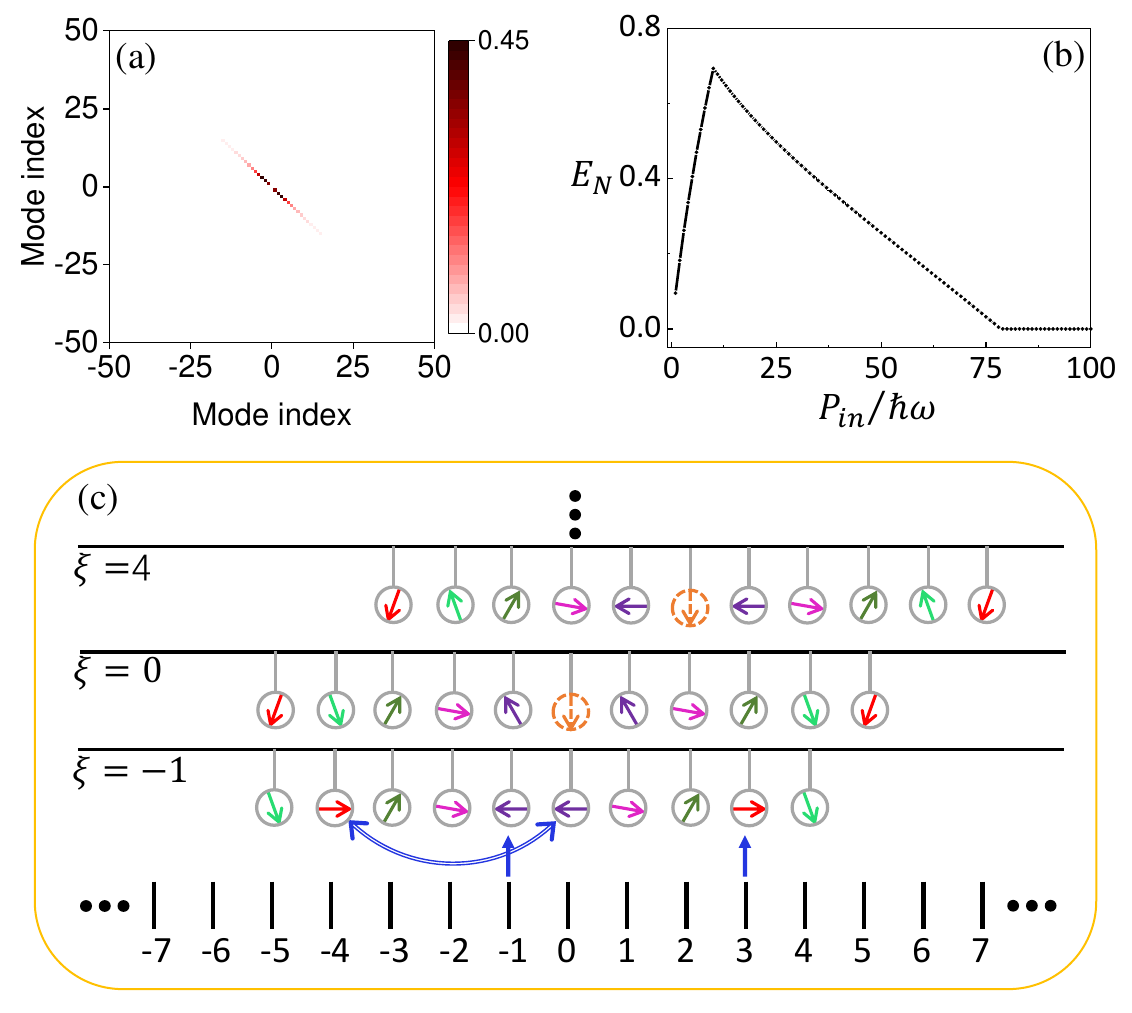}
\par\end{centering}
\caption{Bipartite entanglement in degenerate FWM. (a) Entanglement between different modes for comb below threshold. Only mode pairs symmetric to the pump mode are entangled. (b) Bipartite CV entanglement between the signal and idler modes in a degenerate FWM process. The input photon flux $P_{in}/\hbar\omega$ is normalized to $10^{6}\kappa$. The Kerr coupling rate $g_{3}=10^{-7}\kappa$. (c) FWM processes classified by the conserved total angular momentum
$\xi$, with photon-pairs generated or annihilated simultaneously
in mode-pairs $\left(i,j\right)$ with $\xi=i+j$. The paired modes
are labeled by the same color and orientation, and the index $\xi$
can be taken from $-2N$ to $2N$. Here, we only show $\xi=-1,0,4$
as an illustration. Blue arrow: coherent photon conversion between
modes $m=-4,0$ stimulated by $m=-1,3$, with $\xi=-1$. }
\label{Fig1}
\end{figure}

For all FWM terms in Eq.~(\ref{eq:TotalH-1}), entanglement between two modes is ultimately generated by a photon-pair generation process. By exciting the $0$-th mode with weak pump power, the parametric gain on mode-pair $(i,j)$ with $i+j=0$ cannot compensate for the dissipation in these modes, thus the system stays below the optical parametric oscillation (OPO) threshold and generates thermal photon-pairs by spontaneous parametric downconversion~\citep{reimer2016generation}. In this case, only the $0$-th mode is sufficiently excited, and the intracavity fields in all the other modes are negligible. The effective Hamiltonian reduces to a simple form $H_{\mathrm{eff}}=\sum_{l}g_{\mathrm{eff},l}a_{-l}^{\dagger}a_{l}^{\dagger}$ by neglecting their back action to the pumped mode and FWM among them. Figure~\ref{Fig1}(a) shows the matrix $\mathbf{E}_{N}$, which only has a positive value at its diagonal elements, indicating that only the photon-pair generation between mode-pair $\left(+l,-l\right)$ is initiated. Due to the dispersion of the cavity resonance, modes with indices away from $0$ experience poorer phase-matching conditions and thus $E_{N}$ decays with $l$. This bipartite entanglement can be fully characterized by a degenerate OPO process. For example, Fig.~\ref{Fig1}(b) shows the results of $E_{N}$ for a simple model that only involves a single mode pair at different pump laser powers. By increasing the pump power, the degree of entanglement increases until the system reaches the OPO threshold.

\section{Quantum entanglement in a cascaded FWM network}
It is important to note that the entanglement between mode pairs shows distinct features when the system operates above the OPO threshold, as shown by the sharp downward turn in Fig.~\ref{Fig1}(b). Moreover, it can no longer be described solely by the two-mode squeezing interaction since the generated OPO field with a large amplitude will stimulate other FWM processes. Eventually, a single mode will participate in multiple FWMs and a single FMW also involves multiple optical modes by the cascaded FWM process. These complex interactions between a variety of optical modes in the microring cavity bring DKS unique features in terms of quantum entanglement.

First, with many modes being significantly excited, the number of FWM terms in Eq.~(\ref{eq:TotalH-1}) grows rapidly with the total mode number as $\sim\left(2N+1\right)^{3}$. Since the total angular momentum {[}$\xi=i+j$ of mode-pair $\left(i,j\right)${]} is conserved, we use $\xi$ to classify different FWM terms, as all the mode-pairs of $\xi$ can interact with each other. As shown in Fig.~\ref{Fig1}(c), each horizontal line of $\xi$ represents the group of signal-idler mode-pairs which are labeled by arrows with the same color and orientation. The dashed orange arrow represents the degenerate case $i=j$. It is anticipated that the photon-pair generation for mode-pair $\left(i,j\right)$ would be stimulated by all the coherent light fields belonging to the same $\xi$, which might lead to bipartite CV entanglement by the effective interaction $\left(a_{i}^{\dagger}a_{j}^{\dagger}+a_{i}a_{j}\right)$ for all pairs in $\xi$. Meanwhile, a single mode also connects to multiple $\xi$. As an example, mode $m=-3$ in Fig.~\ref{Fig1}(b) can be entangled with mode $m=3$ with $\xi=0$, $m=2$ with $\xi=-1$, and $m=7$ with $\xi=4$. In addition, for phase-matching condition $i+j=k+l$, there are also coherent photon conversion processes between mode $\left(i,l\right)$ simulated by mode $\left(j,k\right)$ and vice versa, which also distributes the quantum correlations between different modes. Consequently, the FWM in the microresonator results in a complex network, with optical modes serving as nodes and the links corresponding to photon-pair generation and coherent photon conversion. Multicolor quantum entanglement among different moded can be expected from these interactions.

Second, when considering only the second-order dispersion of the cavity, the DKS state has a $\mathrm{sech^{2}}$ shape spectral, with comb lines around the pump mode being excited significantly while modes far wary from the pump stay below the OPO threshold. Therefore, the above analysis based on the two-mode squeezing model applies well to modes far from the pump. However, the DKS state is unique in that only one frequency component is stimulated in each mode near the pump and these frequencies are strictly equally spaced. The self-organization and phase-locking of the comb result in strong suppression of fluctuation and correlation in the combs, which can affect the CV entanglement of the DKS state.

\begin{figure}
\begin{centering}
\includegraphics[width=1\linewidth]{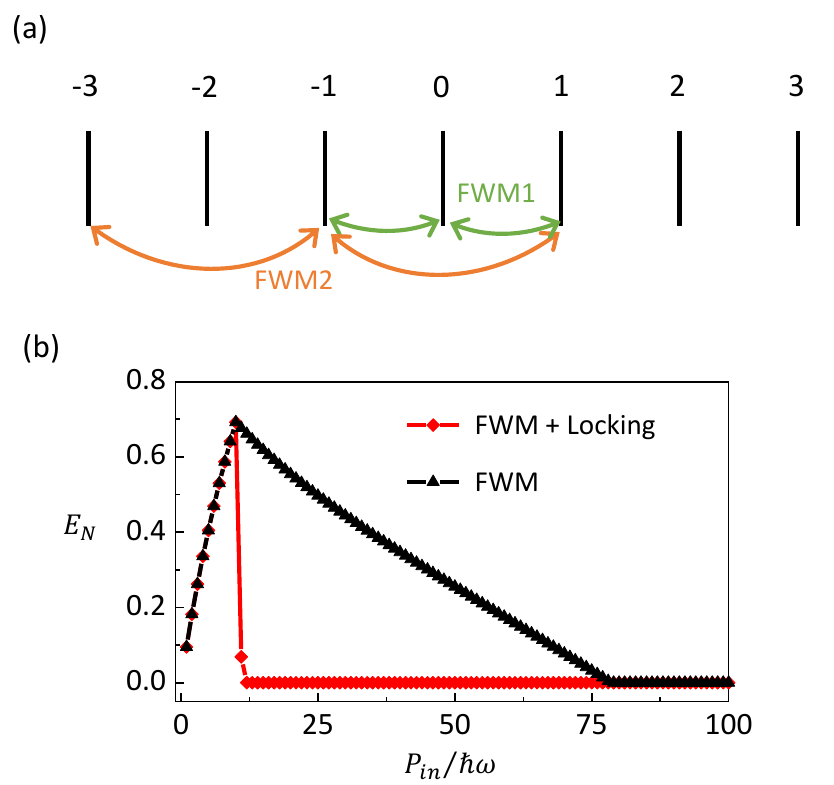}
\par\end{centering}
\caption{Entanglement reduction in the frequency locking process. (a) Illustration of the simplest frequency locking process extracted from the cascaded FWM network for comb generation. Two FWM processes are responsible for the locking process. (b) The CV entanglement between mode-pair $\left\{-1,1\right\}$ for a pure degenerate OPO process (black) and an addition locking interaction (red).}
\label{Fig2}
\end{figure}

In order to obtain a clear physical understanding of the frequency locking mechanism and the associated CV entanglement, we extract the simplest cascaded FWM network that facilitates frequency locking from the full model in Eq.~(\ref{eq:TotalH-1}). As shown in Fig.~\ref{Fig2}(a), it consists of two degenerate FWM processes among the mode family $\left\{-3,-1,0,1\right\}$, which is described by the Hamiltonian
\begin{eqnarray}
H & = & g_{0}a_{0}^{2}a_{-1}^{\dagger}a_{1}^{\dagger}+g_{0}a_{-1}^{\dagger 2}a_{-3}a_{1}+h.c.
\label{eq:FWM-locking}
\end{eqnarray}
The first term is the simple degenerate OPO model that generates pairwise entanglement between mode-pair $\left\{-1,1\right\}$, and the field frequency $\omega_{l}$ in mode $a_{l}$ fulfills $2\omega_{0}=\omega_{-l}+\omega_{l}$. However, the OPO process cannot fully determine the exact value of $\omega_{-1}$, which depends on the mode detuning of the corresponding mode. This means the additional FWM should be introduced to ensure that the comb lines are equally spaced. Here, we introduce a degenerate FWM between mode $a_{-3}$ and modes $a_{-1,1}$ by the second term. This process will generate an additional frequency tone $\omega_{r}=2\omega_{-1}-\omega_{-3}$ in mode $a_{1}$ if $\omega_{-1}-\omega_{3}\neq \omega_{1}-\omega_{-1}$. This frequency tone will serve as an injection seed to compete with the vacuum noise-induced OPO process and eventually lock the laser frequency $\omega_{1}$, i.e., $\omega_{1}$ adjusts itself self-adaptively to match the frequency $\omega_{-3}$ or vice versa. In this way, the fluctuations of the optical fields in both modes $a_{\pm 1}$ and $a_{\pm 3}$ are suppressed. It should be noted that the above analysis is not restricted to the specific mode indices of $\left\{-3,-1,0,1\right\}$, and any mode family that conforms to the Hamiltonian type described in Eq.~(\ref{eq:FWM-locking}) supports the frequency locking. By expanding the five-mode model to encompass the entire mode family, the frequency locking of DKS can be elucidated.

In the comb formation processes, the comb line in mode $a_{-3}$ can be generated either by the degenerate OPO or by other FWM processes, since mode $a_{3}$ is not involved in the locking process. By treating the field in mode $a_{-3}$ as a mean field $\alpha$, an elegant form of the frequency locking Hamiltonian can be approximated from Eq.~(\ref{eq:FWM-locking}) by
\begin{eqnarray}
H & = & g_{0}a_{0}^{2}a_{-1}^{\dagger}a_{1}^{\dagger}+g_{\mathrm{eff}} a_{-1}^{\dagger 2}a_{1}+h.c.
\label{eq:FWM-SHG}
\end{eqnarray}
where the second term describes an effective second-harmonic generation process with $g_{\mathrm{eff}}=g_{0}\alpha$ being the effecting coupling rate. Based on this model we study the change in entanglement associated with the field phase transition of mode $a_{\pm 1}$ during the frequency locking process. 

Figure~\ref{Fig2}(b) depicts the relationship between the degree of entanglement $E_{N}$ of mode-pair $\left\{-1,1\right\}$ and the input photon flux. When the operation is below the OPO threshold, the CV entanglement of the locking process in Eq.~(\ref{eq:FWM-SHG}) (red line) is the same as the bare degenerate OPO (black line) because $\alpha_{-1} \approx 0$ and the system cannot be locked. As the input $P_{in}$ power increases, the system surpasses the threshold and even a small number of photons in mode $a_{-1}$ can lock the state of the system. Here we choose a small coupling rate $g_{\mathrm{eff}}=10^{-7}\kappa$ in the calculation. It can be seen that there is a sudden and sharp drop in the degree of entanglement just after the input exceeds the OPO threshold, which is different from the gradual decline observed in the degenerate OPO process. 

It is worth mentioning that the model in Fig.~\ref{Fig2}(a) exhibits translational symmetry along the mode index. Any mode along with its neighboring modes being excited above the threshold can participate in a locking process. By extending the analysis to stable Kerr comb states, a similar decrease in entanglement can be expected in sufficiently excited comb lines. We also note that there exist various other types of cascaded FWM networks that can result in frequency locking and corresponding entanglement reduction. For instance, the entanglement between mode-pair $\left\{m,n\right\}$ under effective two-mode squeezing interaction $a_{m}^{\dagger}a_{n}^{\dagger}+h.c.$ may also vanish due to the effective interaction $a_{m}a_{n}^\dagger+h.c.$ induced by comb lines in modes $a_{m+l}$ and $a_{n+l}$. These entanglement reduction mechanisms are also applicable to other phased locked states in cascaded nonlinear optics systems.

\section{Entanglement of DKS}
\begin{figure*}
\begin{centering}
\includegraphics[width=16cm]{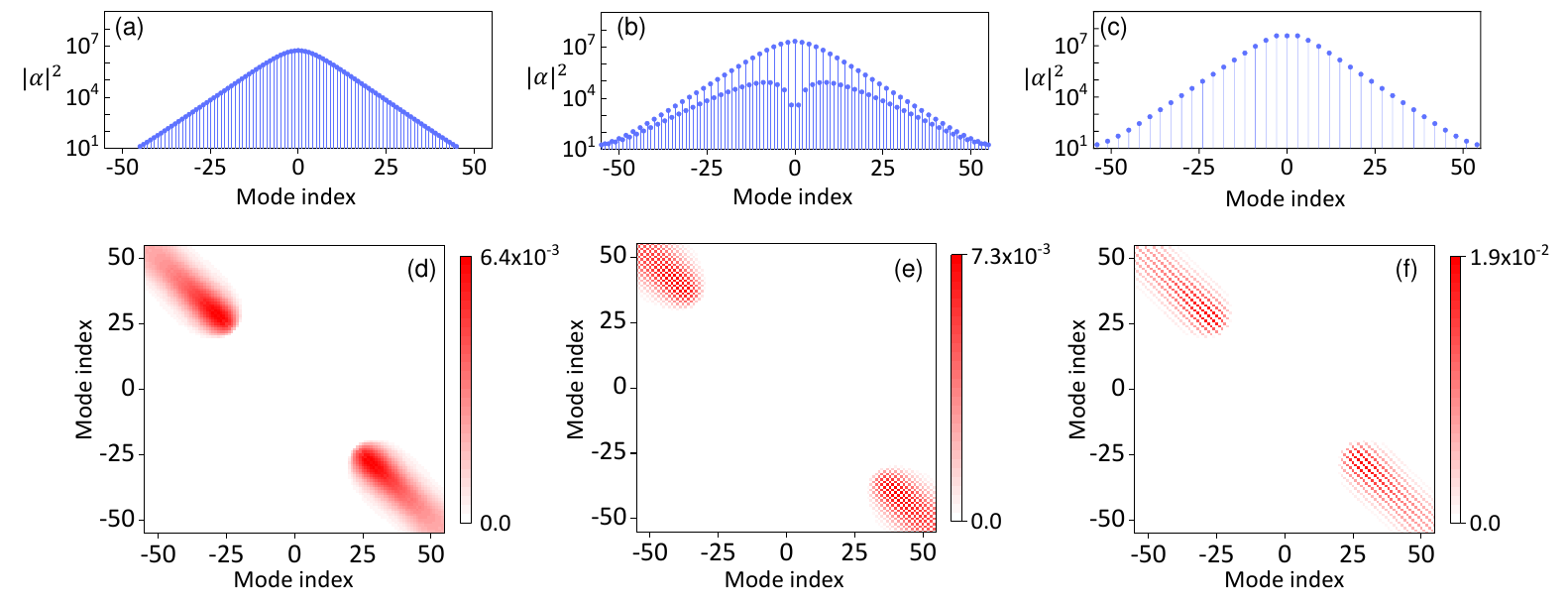}
\par\end{centering}
\caption{Entanglement of different DKS states. (a)-(c) Optical spectral of single-soliton state, two-soliton state, and three-soliton crystal state. (d)-(e) The corresponding $E_{N}$ criterion
of entanglement between all mode-pairs.}
\label{Fig3}
\end{figure*}

Based on the full-model analysis for deriving Eq.(\ref{eq:En}), we numerically calculated the CV entanglement of different DKS states. As the pump power increases above the OPO threshold, the intracavity
field can be driven to the soliton states with appropriate laser power and frequency detuning, which are ultrashort pulses circulating inside the cavity~\citep{herr2014temporal,kippenberg2018dissipative}. Figure~\ref{Fig3} depicts the typical results of Kerr comb spectra in a microring resonator at different DKS
states~\citep{PhysRevA.89.063814,guo2018efficient}, with the upper
and bottom rows shows the classical intracavity fields and the entanglement
$E_{N}$ between mode-pairs. Here, we consider a monochromatic
field driving on the $0$-th mode to initially excite the mode-pairs
with $\xi=0$, and the system parameters are chosen from a typical
AlN microring in the experimental work~\citep{guo2018efficient,Liu2023}.

As shown by the spectra in Figs.\ref{Fig3}(a)-(c), when the cavity is prepared in soliton states of different orders, tens of modes are efficiently excited to equally-spaced comb lines with the space approximately $m\times$FSR determined by the dissipation rate $\kappa$ and the dispersion $D_{2}$~\citep{PhysRevA.89.063814}. The envelopes show a profile of the $\mathrm{sech}^{2}$-function. Even though only $\xi=0$ is initially excited, the generated comb lines can stimulate the cascaded FWM to produce and distribute entanglement for mode-pairs belonging to other $\xi$. The strong comb line with index number $l$ generated by the pump mode can further drive the mode-pairs with $\xi=2l$, leading to the positive $E_{\mathcal{N}}$ on the corresponding diagonals of the entanglement matrix in Fig.~\ref{Fig3}(d)-\ref{Fig3}(f).

It can be directly inferred from Figs.~\ref{Fig3}(d)-(f) that, modes far away from the pump are entangled to several modes on the opposite side of the pump, which is explained by a pure two-body interaction $\left(a_{l}^{\dagger}a_{\xi-l}^{\dagger}+a_{l}a_{\xi-l}\right)$ independently for different $\xi$ and $l$, similar to the case of the below-threshold state {[}Fig.~\ref{Fig1}(a){]}. For the single-soliton state [Fig.~\ref{Fig3}(a)], every mode has a significant comb line. Consequently, $\mathbf{E}_{N}$ in Fig.~\ref{Fig3}(d) shows non-zero values at elements on every diagonal line. For a multi-soliton state, such as the soliton crystal state, the comb lines are most significant for mode index $mn$ with $m$ being the soliton number and $n\in\mathbb{Z}$. According to the intensity distribution of the spectra, the entanglement matrix $\mathbf{E}_{N}$ shows significant positive values only on diagonal lines corresponding to the even mode index ($\xi/2\in\mathbb{Z})$ for two-soliton crystal state and the $3n$ mode index for the three-soliton crystal state [Fig.~\ref{Fig3}(f)]. It should be noted that for non-crystal multi-soliton state, the excitation of other comb lines apart from the mode index $mn$ can also be excited, resulting in more complicated entanglement matrix. Since multi-soliton state has higher energy
than the single-soliton state~\citep{herr2014temporal} under the same pump-cavity detuning, the FWMs
are excited more efficiently, resulting in a higher $E_{N}$. For the case in Fig.~\ref{Fig3}, the there-soliton state has a maximum $E_{N}$ of 0.019, almost three times that of the single-soliton state. 

For all DKS states, we can observe multicolor quantum entanglement by the fact that a single mode is entangled with multiple modes on the opposite side of the comb. In the case of the single-soliton state shown in Fig.~\ref{Fig3}(a), mode $a_{-30}$ is entangled with a total number of $34$ modes ranging from $a_{20}$ to $a_{53}$. As a result of cavity dispersion, modes with a higher mode index have larger detunings, leading to a decrease in both the number of entangled modes and the degree of entanglement. Moreover, the degree of entanglement also decreases as $|\xi|$ increases since the intensity of the comb line decreases with the mode index. 

The DKS has an interesting characteristic that there is no entanglement among modes located near the center of the comb. Since these modes have relatively high intensity, the disappearance of entanglement
can no longer be easily explained by a two-body squeezing interaction. Instead, the cascading of different FWMs forms many frequency locking processes that suppress the quantum fluctuations of the comb lines. The numerical results are in agreement with the simplified locking Hamiltonian model.

\section{Conclusion}
In conclusion, we have investigated the multicolor CV entanglement in the DKS comb and found that the entanglement exhibits distinct characteristics compared to below-threshold combs generated by spontaneous downconversions~\cite{reimer2016generation,Kues2017,Erhard2020,guidry2022quantum}. Our analysis indicates that the reduction of entanglement near the comb center is associated with inherent frequency or phase-locking during DKS formation, which has been analyzed by a simplified model in a complex network of optical modes involving cascaded FWM processes. This mechanism of entanglement reduction can be applied to other frequency-locking processes and phase transition phenomena in nonlinear optics systems. Moreover, we anticipate that by combining with other nonlinear optical effects, the multicolor entanglement of DKS can be further extended to other frequency bands, such as mid-infrared and ultraviolet wavelengths, by $\chi^{(2)}$-nonlinearity~\citep{guo2018efficient,Bruch2021,Liu2023} or Raman scattering~\citep{jung2014green,karpov2016raman,xue2017second}. By multiplexing the color of the DKS to other degrees of freedom of photons, such as the polarization, transverse waveguide mode, and time-bin, a high-dimension quantum cluster state can be generated~\citep{larsen2019deterministic,asavanant2019generation}. Our results reveal the interesting dynamics of quantum entanglement generation associated with DKS and indicate the potential of DKS in quantum information science, such as high-dimensional CV quantum sources, multiparty quantum teleportation networks~\citep{pirandola2015advances}, quantum-enhanced comb sensors, and distributed quantum metrology~\citep{guo2020distributed,xia2020demonstration,zhang2021distributed}.

\smallskip{}

\begin{acknowledgments}
This work was funded by the National Natural Science Foundation of China (Grants 12293053, 92265108, U21A20433, U21A6006, 12104441, 12061131011, and 92265210), and the Natural Science Foundation of Anhui Provincial (Grant No. 2108085MA17 and 2108085MA22). This work was also supported by the Fundamental Research Funds for the Central Universities and USTC Research Funds of the Double First-Class Initiative. The numerical calculations in this paper have been done on the supercomputing system in the Supercomputing Center of the University of Science and Technology of China. This work was partially carried out at the USTC Center for Micro and Nanoscale Research and Fabrication.
\end{acknowledgments}

\end{document}